\documentclass[aps,prb,twocolumn,superscriptaddress,showpacs]{revtex4}
\usepackage{bm}
\usepackage{epsf}
\usepackage{amssymb}
\usepackage{amsmath}
\usepackage{graphicx}
\usepackage{rotating}
\usepackage{epsfig}
\usepackage{epstopdf}
\usepackage{psfrag}
\usepackage{amsmath}
\usepackage{hyperref}
\usepackage{mathrsfs}

\hypersetup{
    bookmarks=true,         
    unicode=false,          
    pdftoolbar=true,        
    pdfmenubar=true,        
    pdffitwindow=true,      
    pdftitle={My title},    
    pdfauthor={Author},     
    pdfsubject={Subject},   
    pdfcreator={Creator},   
    pdfproducer={Producer}, 
    pdfkeywords={keywords}, 
    pdfnewwindow=true,      
    colorlinks=true,       
    linkcolor=red,          
    citecolor=blue,        
    filecolor=magenta,      
    urlcolor=cyan           
}

\DeclareMathAlphabet{\bi}{OML}{cmm}{b}{it}

\begin{document}
\def\G{{\cal G}}
\def\F{{\cal F}}
\def\ea{\textit{et al.}}
\def\bM{{\mathbf M}}
\def\bM{{\mathbf M}}
\def\bU{\hat{{\bm U}}}
\def\bG{\hat{{\bm G}}}
\def\bGt{\hat{\tilde{{\bm G}}}}
\def\bV{{\bm V}}
\def\bj{\bm{j}}
\def\bSig{{\bm \Sigma}}
\def\bLam{{\bm \Lambda}}
\def\bDel{{\vec{\mathbf{\Delta}}}}
\def\bn{{\bf n}}
\def\d{{\bf d}}
\def \xy{$x$--$y$ }
\def\bP{{\bf P}}
\def\bK{{\bf K}}
\def\bk{{\bf k}}
\def\bkn{{\bf k}_{0}}
\def\bz{{\bf z}}
\def\bR{{\bf R}}
\def\br{{\bf r}}
\def\bx{\mathbf{\hat{x}}}
\def\by{\mathbf{\hat{y}}}
\def\bq{{\bf q}}
\def\bp{{\bf p}}
\def\bD{{\hat{\mathbf{D}}}}
\def\bQ{{\bf Q}}
\def\bs{{\bf s}}
\def\bA{{\mathbf A}}
\def\bv{{\bf v}}
\def\b0{{\bf 0}}
\def\la{\langle}
\def\ra{\rangle}
\def\Im{\mathrm {Im}\;}
\def\Re{\mathrm {Re}\;}
\def\bea{\begin{align}}
\def\eea{\end{align}}
\def\beq{\begin{equation}}
\def\eeq{\end{equation}}
\def\bdm{\begin{displaymath}}
\def\edm{\end{displaymath}}
\def\bnab{{\bm \nabla}}
\def\Tr{{\mathrm{Tr}}}
\def\bJ{{\bf J}}
\def\bU{{\bf U}}
\def\bPsi{{\bm \deltaDelta}}
\def\mA {\mathrm{A}}
\def \R{R_{\mathrm{s}}}
\def \rhos{n_{\mathrm{s}}}
\def \rhon{\tilde{n}}
\def \Rd{R_{\mathrm{d}}}
\def \xy{three dimensional $XY\;$}
\def\sfrac{\textstyle\frac}
\def\e0{\epsilon_B}
\def\ath{a_{3d}}
\def\atw{a_{2d}}
\def\ntw{n_{\mathrm{2D}}}
\def\2d{2d}
\def\3d{3d}
\def\g2{\ln(k_F\atw)}
\def\bSig{\hat{\mathbf{\Sigma}}}
\def\bGama{\hat{\mathbf{\Gamma}}_{\alpha}}
\def\bGamn{\hat{\mathbf{\Gamma}}_{0}}
\def\bGammu{\hat{\mathbf{\Gamma}}_{\mu}}
\def\bFmu{\hat{\mathbf{F}}_{n\mu}}
\def\bF0{\hat{\mathbf{F}}_{n 0}}
\def\bgammu{\hat{\mathbf{\gamma}}_{\mu}}
\def\bGamb{\hat{\mathbf{\Gamma}}_{\beta}}
\def\bGamx{\hat{\mathbf{\Gamma}}_{x}}
\def\bGamy{\hat{\mathbf{\Gamma}}_{y}}
\def\bgama{\hat{\mathbf{\gamma}}_{\alpha}}
\def\bgamb{\hat{\mathbf{\gamma}}_{\beta}}
\def\bgamx{\hat{\mathbf{\gamma}}_{x}}
\def\bgamy{\hat{\mathbf{\gamma}}_{y}}
\def\bgamxt{\hat{\tilde{\mathbf{\gamma}}}_{x}}
\def\bgamyt{\hat{\tilde{\mathbf{\gamma}}}_{y}}
\def\Sr{\mathrm{Sr}_2\mathrm{RuO}_4}
\def\sigI{\sigma^{\mathrm{(I)}}_H(\nu)}
\def\sigII{\sigma^{\mathrm{(II)}}_H(\nu)}
\def\sigIdc{\sigma^{\mathrm{(I)}}_H(0)}
\def\sigIIdc{\sigma^{\mathrm{(II)}}_H(0)}
\def\wh{\widehat}
\def\G{{\mathcal G}}
\def\F{{\mathcal F}}

\def\A{{A}}

\def\gan{\hat{\mathbf{\gamma}}_{\alpha}}
\def\gaf{\hat{\mathbf{\Gamma}}_{\alpha}}
\def\gbn{\hat{\mathbf{\gamma}}_{\beta}}
\def\gbf{\hat{\mathbf{\Gamma}}_{\beta}}
\def\gmn{\hat{\mathbf{\gamma}}_{\mu}}
\def\gmf{\hat{\mathbf{\Gamma}}_{\mu}}
\def\gof{\hat{\mathbf{\Gamma}}_{0}}

\title{Non-topological nature of the edge current in a chiral $p$-wave superconductor}
\author{Wen~Huang}
\affiliation{Department of Physics and Astronomy, McMaster University, Hamilton, Ontario, L8S 4M1, Canada}
\author{Samuel Lederer} 
\affiliation{Department of Physics, Stanford University, Stanford, California, 94305, USA}
\author{Edward~Taylor}
\affiliation{Department of Physics and Astronomy, McMaster University, Hamilton, Ontario, L8S 4M1, Canada}
\author{Catherine~Kallin}
\affiliation{Department of Physics and Astronomy, McMaster University, Hamilton, Ontario, L8S 4M1, Canada}
\affiliation{Canadian Institute for Advanced Research, Toronto, Ontario M5G 1Z8, Canada}

\date{Dec. 16, 2014}

\begin{abstract}
The edges of time reversal symmetry breaking topological superconductors support chiral Majorana bound states as well as spontaneous charge currents. The Majorana modes are a robust, topological property, but the charge currents are non-topological--and therefore sensitive to microscopic details--even if we neglect Meissner screening. We give insight into the non-topological nature of edge currents in chiral p-wave superconductors using a variety of theoretical techniques, including lattice Bogoliubov-de Gennes equations, the quasiclassical approximation, and the gradient expansion, and describe those special cases where edge currents do have a topological character. While edge currents are not quantized, they are generically large, but can be substantially reduced for a sufficiently anisotropic gap function, a scenario of possible relevance for the putative chiral p-wave superconductor $\Sr$. 

\end{abstract}

\maketitle

\section{Introduction}

Time reversal symmetry breaking topological superconductors support branches of chiral Majorana bound states at their edges~\cite{Beenakker13}.  The number of these branches is insensitive to perturbations such as weak disorder, and is equal to a Chern number, a topological invariant that is determined by the Fermi surface topology and the chirality of the order parameter.   Quantum Hall systems support both topologically protected edge states and topologically protected, \emph{quantized} edge currents, with the conductance equal to fundamental constants multiplied by a Chern number~\cite{Niu85}.  Even though the number of topological edge modes is given by a Chern number in both these systems, the edge current of a topological superconductor is not topologically protected or quantized.  This fact is clear from Bogoliubov-de Gennes (BdG) calculations~\cite{Imai12,Imai13,Lederer14} of topological chiral $p$-wave superconductors that reveal non-universal behaviour dependent on microscopic details. This non-universal behaviour is present even without taking into account the effects of Meissner screening (which we neglect here), which forces the total current to vanish (though the local currents should still yield  observable magnetic signals \cite{Matsumoto99,Furusaki01}).  One reason to study this issue is that the lack of topological protection of edge currents in chiral $p$-wave superconductors is crucial to any attempt to reconcile the null result of precision magnetometry experiments on the putative chiral $p$-wave superconductor $\Sr$\cite{Kirtley07,Hicks10,Jang11,Curran14} with straightforward theoretical predictions\cite{Matsumoto99}.

In this work we provide insight into the non-topological nature of edge currents in chiral $p$-wave superconductors using a variety of theoretical techniques, including lattice Bogoliubov-de Gennes equations, the quasiclassical approximation, a gradient expansion of the effective action, and spectral flow arguments. We begin by examining the circumstances under which topology {\it does} straightforwardly govern edge currents: (i) the coupling is weak, so that $\Delta_0\ll E_F$ (assumed throughout this paper) and (ii) a spatially varying site energy $A_0(\br)$ (equivalent to a static, unscreened scalar potential)  drives the density to zero at the edge over a distance $L$ much longer than the coherence length $\xi_0$. (We will refer to condition (ii) as the {\it soft edge} limit). Under these circumstances the gradient expansion gives\cite{Volovik88,Volovik92,Goryo98,Stone04}
\beq \bj(\br) = -\frac{C}{4\pi}(\hat{\mathbf{z}}\times \bnab)A_0(\br) 
\label{JH0}\eeq
for the current density where $C$ is the Chern number and we use units where the electron charge $e=\hbar=1$ throughout.  Apart from a factor of one-half, this is also the current density in quantum Hall systems, both in fractional quantum Hall systems where the Chern--Simons action was first derived in a condensed matter context~\cite{Girvin87,Zhang89}, as well as integer quantum Hall systems. In the quantum Hall context, \eqref{JH0} implies a quantized, topological value for the Hall conductance.

Contrary to the assumptions above, the edges of actual superconducting crystals are atomically sharp: the density at the edge vanishes over an atomic scale $k^{-1}_F\ll \xi_0$.  This explicitly invalidates the systematic gradient expansion in powers of $\xi_0/L$. Even within the gradient expansion there are subleading corrections to (\ref{JH0}) whose importance grows as $L$ is diminished; one such correction is discussed in Sec.~\ref{topologysec}.  That said, despite the fact that \eqref{JH0} fails to even approximately describe the current {\it density} in the sharp edge limit, there are special models with sharp edges for which the {\it integrated} current (which is roughly proportional to the strength of the magnetic signal expected in experiment) coincides with the prediction of \eqref{JH0}. These special models include all continuum models (for which the integrated current can be calculated using a one-dimensional Dirac equation~\cite{Stone04,Sauls11}), as well as certain lattice models with restricted hopping matrix elements. We analyze these special models in Sec.~\ref{nontoposec}, using the ``spectral flow''~\cite{Volovik95} properties of the BdG eigenvalues to show that the integrated current remains equal to its ``topological value" (i.e. the one inferred from the Chern--Simons expression (\ref{JH0})) as the edge is deformed from soft to sharp.

Outside of these special models, the integrated edge current generically evolves to a non-topological value (i.e. one unrelated to the Chern number)  as we adiabatically deform a soft edge into a sharp one. 
While it remains generically substantial, there is nothing to prevent it from being small, and it can be tuned through zero by varying the band and/or gap-structure.  For example, in a model with an anisotropic $p$-wave order parameter consistent with
 next-nearest-neighbor (NNN) pairing \cite{Scaffidi14} on the $\gamma$ band of $\Sr$, 
 the integrated edge current vanishes at a filling fraction close to  the experimental value (see Fig. \ref{current2fig}).
 Although reliant on fine-tuning of parameters, this result might be important for reconciling chiral $p$-wave superconductivity in $\Sr$~\cite{Mackenzie03,Kallin09,Kallin12,Maeno12} with the null results of experiments designed to measure the expected magnetic fields \cite{Kirtley07,Hicks10,Jang11,Curran14}.

\section{Topological properties in the continuum limit}
\label{topologyHe3sec}

The topological properties of a two-dimensional chiral $p$-wave superfluid are characterized by the Chern number 
\begin{align}
C =\frac{1}{4\pi}\int d^2k\; \hat{h} \cdot \left( \partial_{k_x} \hat{h} \times \partial_{k_y} \hat{h}\right).\label{Chern}\end{align}
Here $\vec h =\left\{\mathrm{Re}[\Delta_0(\bk)],-\mathrm{Im}[\Delta_0(\bk)],\xi(\bk)\right\}$ and $\hat h = \vec h/|\vec h|$.  $\Delta_0(\bk)$ is the complex chiral order parameter and $\xi(\bk) \equiv \epsilon(\bk)-\mu$, with $\epsilon(\bk)$ the single-particle dispersion.  For a chiral $p$-wave order parameter $\Delta_0(\bk) = \Delta_0(k_x\pm ik_y)/k_F$ appropriate for continuum systems, the Chern number is  $\pm 1$.   For lattice models, it depends not just on the chirality or winding of the order parameter, but also the topology of the Fermi surface, but  always takes an integer value.  

One manifestation of a non-zero Chern number is a quantized value of the ``static'' Hall conductivity~\cite{Volovik88,Volovik92,Goryo98,Stone04,Roy08}: $\tilde{\sigma}_{xy}\equiv \lim_{\bq\to 0}\lim_{\omega\to 0}\sigma_{xy}(\omega,\bq) = C/4\pi + {\cal{O}}[(\Delta_0/E_F)^2]$ in the weak-coupling limit, a result that follows from (\ref{JH0}) (which we derive in Sec.~\ref{topologysec}).  Note that in a continuum system, reversing the order of limits to evaluate the standard DC Hall conductivity $\sigma_{xy} \equiv \lim_{\omega\to 0}\lim_{\bq\to 0}\sigma_{xy}(\bq,\omega)$, gives zero~\cite{Roy08}.  This non-commutativity of limits arises from a subtlety in the effective action (\ref{Leff}) which we will discuss later on.

Closely related to this static quantum Hall effect is the fact that a long-wavelength density perturbation of a chiral $p$-wave superfluid will give rise to a quantized current.  With $C=1$ and $A_0(\br)$ determining the local carrier density $n(\br)$ according to $\nabla A_0(\br)=\pi \nabla n(\br)/m$
(\ref{JH0}) reduces to the well-known expression for the current in a chiral superfluid due to Mermin and Muzikar~\cite{Mermin80}:
 \beq \bj = \frac{1}{4m}\left({\bf \hat z} \times \nabla n  \right).\label{MMcurrent}\eeq

Using this result to evaluate the edge current, assuming that the density evolves \emph{slowly} 
from zero at $x=-\infty$ to its bulk value $n_0$ at $x=+\infty$, the integrated current is 
\beq I_{y} = \frac{C}{4m}\int^{-\infty}_{-\infty} dx \partial_x n(x) = \frac{n_0C}{4m}.\label{Iysigma}\eeq
Remarkably, this result agrees with calculations of the edge current in a Galilean invariant chiral $p$-wave superfluid by Stone and Roy~\cite{Stone04} (using BdG), and Sauls~\cite{Sauls11} (using the quasiclassical approximation).   This is surprising because these results are obtained for a sharp edge whereas (\ref{JH0}) is obtained from a gradient expansion of the action and should only be strictly valid in the soft-edge limit.  Despite this, both Volovik~\cite{Volovik92} and Goryo and Ishikawa~\cite{Goryo98} have used this result to conclude that the edge current in a Galilean-invariant chiral $p$-wave superfluid is quantized.

\section{Model and BdG results}
\label{preliminarysec}

We now turn to BdG calculations of the edge current for a range of one-band lattice models of chiral $p$-wave superconductivity.  For simplicity, we consider spinless fermions on a two-dimensional square lattice (we will multiply our results for the current by two to compensate):
\beq
H=-\sum_{\br,\br'} t_{\br,\br'}c^{\dagger}_{\br}c_{\br'}- \mu\sum_{\br} c^{\dagger}_{\br}c_{\br} -\sum_{\br,\br'}g_{\br,\br'}c^{\dagger}_{\br}c^{\dagger}_{\br'}c_{\br'}c_{\br}. \label{H}
\eeq
Here $\br,\br'$ denote the lattice positions,  $t \equiv t_{\br,\br \pm \bx}=t_{\br,\br \pm \by}$ and $t^{\prime}\equiv t_{\br,\br\pm(\bx \pm \by)}$ are the nearest- (NN) and next-nearest-neighbour (NNN) hopping parameters.  
Decoupling the interaction term by introducing the two-component order parameter $(\Delta_x,\Delta_y)$, the pairing term in the Hamiltonian is 
\begin{align} H_{\Delta} =&\sum_{\br,\bs}\left[\Delta_x(\br,\bs)+\Delta_y(\br,\bs)\right]c^{\dagger}_{\br-\bs/2}c^{\dagger}_{\br+\bs/2}+\mathrm{H.c.}\label{Hpair}\end{align}
We select the chiral $p$-wave channel by taking a relative phase of $\pi/2$ between $\Delta_x$ and $\Delta_y$, and by assuming $\Delta_x$ and $\Delta_y$ transform, respectively, under the $p_x$ and $p_y$ representation of the square lattice point group.  Assuming that pairing occurs in a single lattice harmonic, $\Delta_{\alpha}(\br,\bs) \equiv \eta_{\alpha}(\br)\Delta_{0,\alpha}(\bs)$ can be written as a separable function of the centre-of-mass $\br$ and relative $\bs$ coordinates, where $\eta_{\alpha}(\br)$ is the dimensionless amplitude, equal to unity in the bulk.  
As is well-known, this model supports chiral Majorana modes at the edges of the superconductor.  Modulo a sign factor, the number of such chiral modes per edge is given by (\ref{Chern}), where now 
\beq \xi(\bk) = -2t(\cos k_x + \cos k_y) - 4t'\cos k_x\cos k_y - \mu,\label{xi}\eeq
and
and
\beq \Delta_0(\bk) = \Delta_{0,x}(\bk)+\Delta_{0,y}(\bk) \label{OP}\eeq
where $\Delta_{0,\alpha}(\bk)$ is the Fourier transform of $\Delta_{0,\alpha}(\bs)$. For the simplest case of NN pairing, $\Delta_0(\bk)= \Delta_0(\sin k_x \pm i\sin k_y).$

 \begin{figure}[!hr]
\centering
\begin{minipage}[r]{0.33\textwidth}
\includegraphics[width=1.0\linewidth]{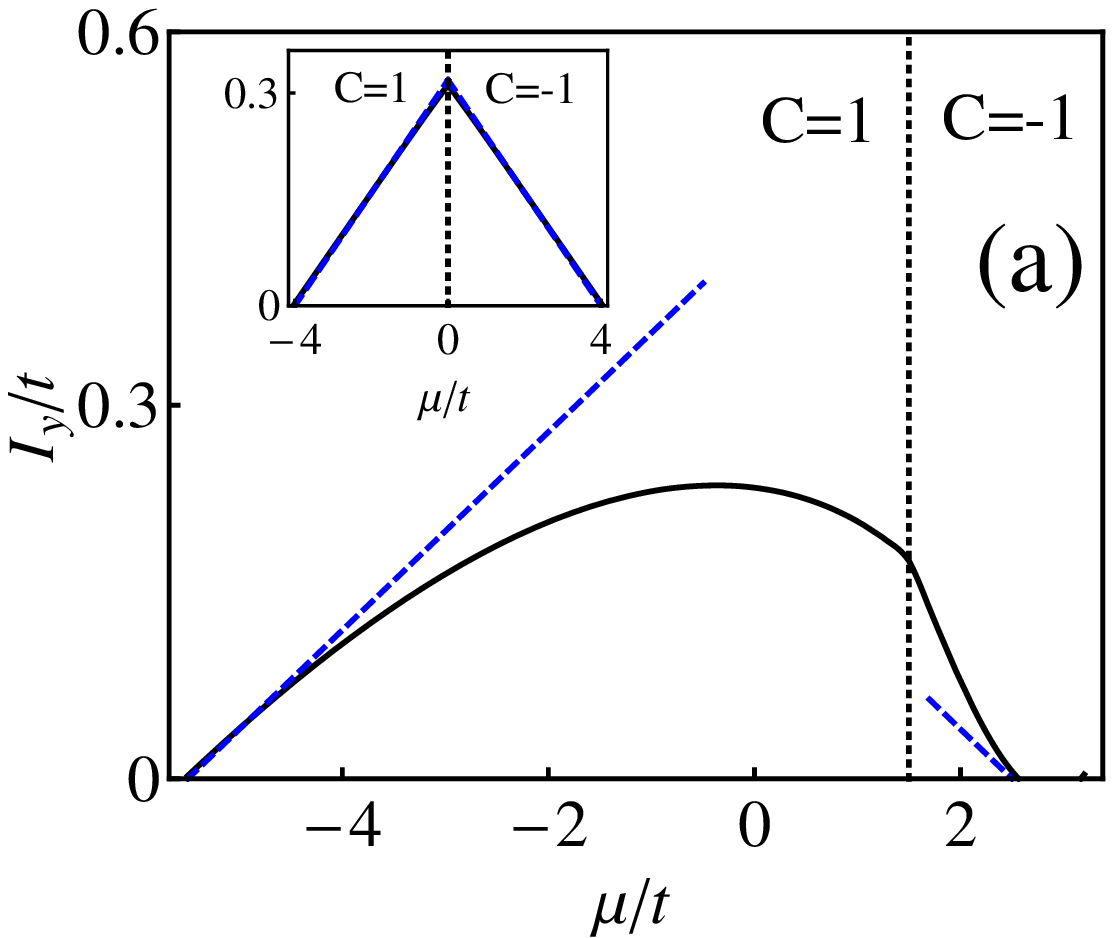}
 \vspace{0.0cm}
\end{minipage}
\begin{minipage}[r]{0.35\textwidth}
\includegraphics[width=1.0\linewidth]{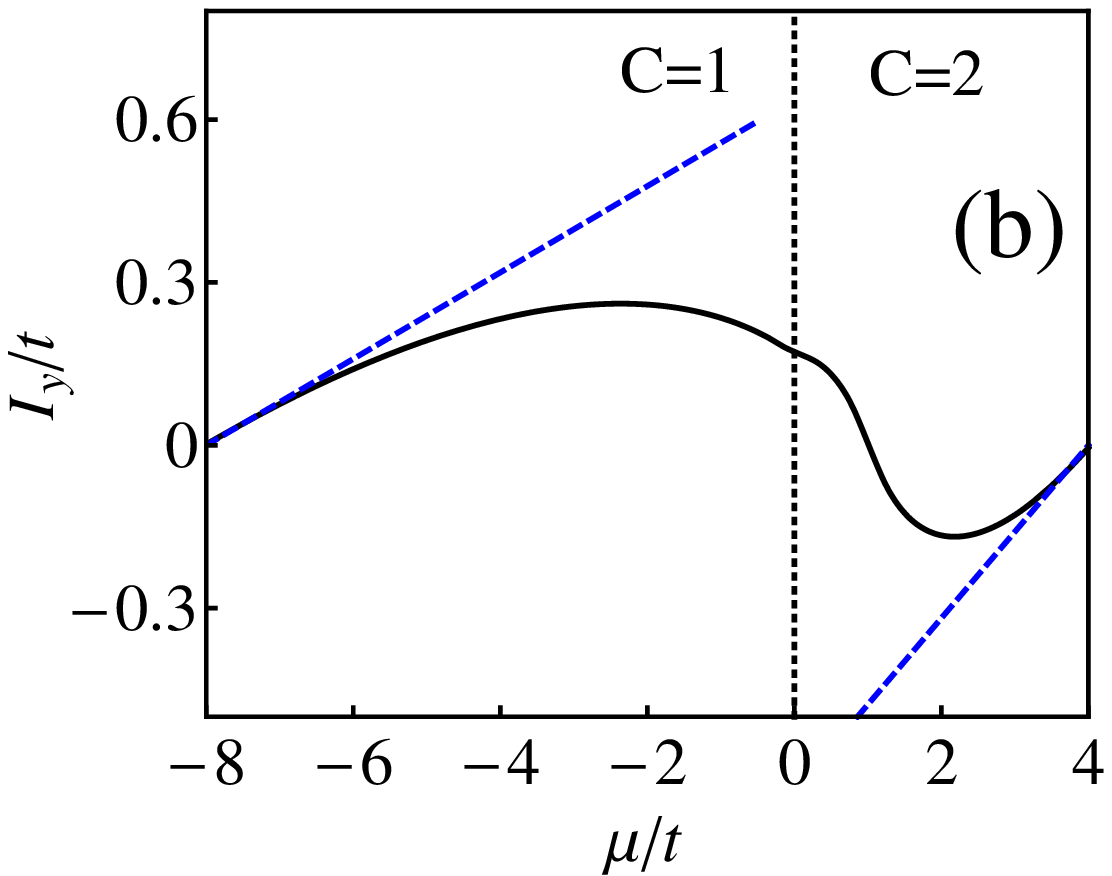}
\end{minipage}
\caption{The integrated edge current calculated from $T=0$ BdG (solid curves) as a function of the chemical potential over the entire bandwidth for: (a) $t'=(3/8)t$ [inset: $t'=0$] and (b) $t'=t$.  $\Delta_0 = 0.2t$ for all plots; this requires varying the interaction $g$ as the chemical potential is varied.   In the inset of (a) for instance, $g/t$ is varied from 11.2 at $\mu= -4t$ to 3.25 at $\mu=0$.  Calculations are carried out for $N_x=N_y=300$ lattice sites. The ``topological current" obtained from \ref{JH0}, with details given in Appendix~\ref{topcurrentapp}, is also shown (dashed lines) and coincides with the BdG result for $t'=0$. Regions of $\mu$ with different Chern numbers are separated by a dotted vertical line.}
\label{currentfig}
\end{figure}

To calculate the edge current in this model, we perform BdG calculations in a cylindrical geometry: periodic boundary conditions are taken in the $y$ direction, and open boundary conditions in $x$.   The current operator for the link from site $i$ to site $j$ is 
\begin{equation}
\hat J_{i,j}=i d_{ij} t_{i,j} \left[c^{\dagger}_{i}c_{j}-c^{\dagger}_{j}c_{i,}\right],
\end{equation}
where $d_{ij}$ is the bond length connecting $i$ and $j$.  (Here, in addition to $\hbar=1$, we set $e=1$; the unit cell length $a$ is also set to unity so that $d_{ij}=1$ for NN sites and $d_{ij}=\sqrt{2}$ for NNN sites.)  
Our primary results for the edge currents will concern the total current $I_y$ flowing through one half of the cylindrical system along the $y$-direction.  Let the cylinder be $N_x$ sites wide, and $N_y$ sites in circumference (in all our calculations, we use $N_x=N_y$).  Then
\begin{equation}
I_y=\sum^{N_x/2}_{n=1} \langle \hat J_{n\hat x, n\hat x+\hat y}+\frac{1}{\sqrt{2}}\hat{J}_{n\hat x, n\hat x+\hat x+ \hat y}\rangle, \label{Iy0}
\end{equation}
where the two terms in the sum are for NN and NNN links.  

\begin{center}
\begin{figure}
\includegraphics[width=0.8\linewidth]{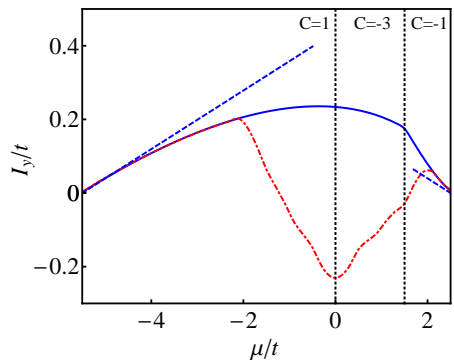}
\caption{Effect of order parameter anisotropy on the edge current.  The integrated edge current for two different order parameters is shown for $t'=3t/8$: $\Delta_0(\bk)=\Delta_0(\sin k_x + i\sin k_y)$ [solid curve; same as in Fig~\ref{currentfig}(a)] and $\Delta_0(\bk) = \Delta_0(\sin k_x\cos k_y + i\sin k_y \cos k_x)$ (red dot-dashed curve). The Chern number in the latter case is equal to 1 for $-5.5t\leq \mu< 0$, -3 for $0< \mu<1.5t$, and -1 for $1.5t<\mu< 2.5t$.  The topological current value is shown by the blue dashed curves and coincides for the two order parameters.}  
\label{current2fig}
\end{figure}
\end{center}

The  mean-field Hamiltonian comprised of the single-particle terms of (\ref{H}) and the pairing contribution (\ref{Hpair}) is diagonalized, and self-consistency is enforced by iterating the gap equation.  Figure~\ref{currentfig} shows the integrated edge current as a function of the chemical potential for different values of the NNN hopping $t'$.  We also show the ``topological" (soft-edge limit) expressions for the integrated currents near the continuum limit at the top and bottom of the band obtained from (\ref{JH0}) using the Chern numbers for these models (although this topological expression is not always uniquely defined, as we discuss in Appendix~\ref{topcurrentapp}).  Apart from a coincidence in the case where $t'=0$ [shown in the inset of Fig.\ref{currentfig} and elaborated on in Sec.~\ref{nontoposec}], these topological values differ considerably from the BdG results, explicitly demonstrating the non-topological nature of the edge current.   

Figure~\ref{current2fig}  compares the integrated current as a function of chemical potential for models with chiral $p$-wave order parameters of dramatically different anisotropy: the NN pairing case, $\Delta_0(\bk) = \Delta_0(\sin k_x +i\sin k_y)$, and  the NNN pairing case, $\Delta_0(\bk) = \Delta_0(\sin k_x \cos k_y + i\sin k_y \cos k_x)$.  For comparison, we also show a suitably defined topological expression which, away from the top and bottom of the band, fails to even qualitatively track the current in the case of NNN pairing.  For parameters appropriate for $\Sr$, $\mu\sim 1.4t$~\cite{Damascelli00}, the current obtained for NNN pairing is significantly reduced compared to that for NN pairing.  

All our numerical BdG results are well approximated by the expression 
\beq I_y=\frac{1}{(2\pi)^d}\oint_{FS} \frac{d^{d-1}\bk}{|{\bf v}|}v_x v_y\tan^{-1}\left(\frac{\Delta_{0,x}}{\Delta_{0,y}}\right).\label{ISC}\eeq
which is derived in detail using the quasiclassical approximation in Appendix~\ref{QCapp}.
Here the subscript ``$FS$'' denotes an integral over the Fermi surface, $v_{\mu}\equiv \partial_{\bk_{\mu}}\xi(\bk)$, $| \bv | \equiv \sqrt{v_x^2+v_y^2}$, and $\Delta_{0,x}(\bk)$ and $\Delta_{0,y}(\bk)$ are the momentum dependent order parameter components [c.f. (\ref{OP})].  This result confirms that the edge current is generically equal to the Fermi energy times a number of order one and fundamental constants. However, there can be substantial cancellations in the integral of \eqref{ISC} if the order parameter components have ``accidental'' sign changes around the Fermi surface, as occurs for the anisotropic gap shown in Fig.~\ref{current2fig} at sufficiently large carrier density.  For certain non-$p$-wave chiral order parameters such as $d_{xy}+id_{x^2-y^2}$ on a square lattice, symmetry requires this cancellation to be complete and the current vanishes identically within a quasiclassical approximation~\cite{Huang14,Tada14}.

\section{Gradient expansion of the BCS action for a chiral $p$-wave superconductor}
\label{topologysec}

To complement our BdG results, we now turn to a gradient expansion of the mean-field BCS action for a chiral $p$-wave superconductor.   Previous authors~\cite{Volovik92,Goryo98,Stone04} have used such an expansion of the action with respect to gradients of the scalar $A_0(\br)$ potential to understand the edge current.  A vector potential $\bA(\br)$ is also included to generate an expression for the current from the action, taking it to be zero after this is done.  At $T=0$, and in the weak-coupling limit, the leading-order terms that give rise to a spontaneous current in this gradient expansion are  (see Appendix~\ref{actionapp})
\beq {\cal{L}}_{\mathrm{eff}} = 
-\frac{C}{4\pi}\epsilon_{0\mu\nu}\A_{\mu}\partial_{\nu}\A_{0},\label{Leff}\eeq
where implicit summation over the Cartesian indices $\mu,\nu$ is assumed.  $C$ is the Chern number defined in (\ref{Chern}), and $\epsilon_{\lambda\mu\nu}$ is the Levi-Civita symbol corresponding to space-time indices $(0,1,2)=(\tau,x,y)$.  (\ref{Leff}) resembles the Chern--Simons term which arises in the effective theory of the fractional quantum Hall effect~\cite{Girvin87,Zhang89}.   Unlike in that theory, the ``Chern--Simons-like'' action (\ref{Leff}) lacks the time derivative $-(C/4\pi)\epsilon_{\mu0\nu}\A_{\mu}\partial_{0}\A_{\nu}$~\cite{Stone04}.  The absence of this only affects dynamic properties such as the Hall response discussed earlier and not static ones such as the edge current, and hence, is not responsible for the non-topological nature of the edge current.  Applying $\bj = \left.\delta{\cal{L}}_{\mathrm{eff}}/\delta \bA\right|_{\bA=0}$ to (\ref{Leff}) gives the result (\ref{JH0}) for the current.

The gradient expansion leading to (\ref{Leff}) is strictly valid only when $A_0(\br)$ varies on length scales much longer than the superconducting coherence length.  This is the opposite limit to a sharp crystalline edge, where the density varies over an atomic scale $k^{-1}_F\ll \xi_0$, so the gradient expansion formally breaks down.  As one moves away from the soft-edge limit, there will be gradient corrections involving $A_0$ beyond (\ref{Leff}).  In addition, one expects the order parameter amplitudes $\eta_x$ and $\eta_y$ to vary in space differently in response to the presence of an edge~\cite{Ambegaokar74}.  Including such textures in the gradient expansion of the BCS action leads to a term
\beq {\cal{L}}_{\Delta} = \gamma\left[\A_y\partial_x +\A_x\partial_y\right](\eta_y-\eta_x),\label{LD}\eeq
where
\beq
 \gamma \equiv \int \frac{d^2\bk}{(2\pi)^2} \frac{v_xv_y\mathrm{Im}(\Delta^{*}_{0,x}(\bk)\Delta_{0,y}(\bk))}{2E^3_{\bk}},\label{gammaxy}\eeq
with $E_{\bk}=\sqrt{\xi^2(\bk)+|\Delta_0(\bk)|^2}$. (\ref{LD}) gives rise to an additional, non-topological contribution
\beq j_{\Delta,i}(\br) \equiv \gamma\epsilon_{ij} \partial_j[\eta_i(\br)-\eta_j(\br)],\label{JD}\eeq
to the edge current.

(\ref{JD}) is the zero-temperature analogue of the usual Ginzburg--Landau expression (see e.g., Ref.~\onlinecite{Ashby09}) for the current in the absence of an explicit potential $A_0(\br)$.  (For $A_0(\br)\neq 0$, there is also an analogue of the Chern--Simons term at $T=T_c$~\cite{Huang14}.)  While the expansion involving gradients of $A_0$ breaks down completely in the sharp-edge limit~\cite{gradientnote}, (\ref{LD}) remains qualitatively valid since the order parameter textures vary over the coherence length, putting this term at the edge of the domain of validity of our gradient expansion.  The same calculation that yields $\gamma$ at $T=0$ gives the GL coefficient $k_3=k_4$ at $T\simeq T_c$~\cite{Huang14}.  At $T=0$, in the continuum limit, it reduces to $\gamma = \mu/8\pi \simeq n/8m$, showing that this contribution to the edge current is generically substantial.  Indeed, the $\gamma$ coefficient bears a qualitative resemblance to the quasiclassical expression (\ref{ISC}) for the current.  Calculating the integrated current that results from  (\ref{JD}) using (\ref{gammaxy}) and self-consistent values of $\eta_x(\br), \eta_y(\br)$ from BdG calculations, the result is in qualitative agreement with numerical BdG for all lattice structures and gap anisotropies studied.

\section{Topological and non-topological aspects of the edge current}
\label{nontoposec}

\begin{center}
\begin{figure}
\includegraphics[width=0.7\linewidth]{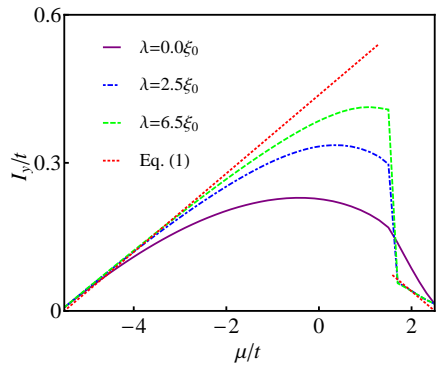}
\caption{Plots of the integrated edge current from BdG for $t'=3t/8$ [see also Fig.~\ref{currentfig}(b)] with an edge at $x=0$ and an edge potential $A_0(x) = (\mu+5.5t) (1-\tanh(x/\lambda))$ for $\mu<1.5t$ and $A_0(x) = (\mu-2.5t) (1-\tanh(x/\lambda))$ for $\mu>1.5t$.  As the edge becomes progressively softer ($\lambda/\xi_0$ increasing), the BdG  results approach the topological value (\ref{Iy2}) obtained from (\ref{JH0}).  All results should coincide near the bottom ($\mu=-5.5t$) and top ($\mu=2.5t$) of the band.  The van Hove singularity at $\mu=1.5t$ pushes the region of agreement near the top of the band to values of $\mu$ very close to $2.5t$.  For $\lambda=6.5\xi_0$, the current does not vanish at the top of the band since we had to use a large value of the order parameter, $\Delta_0=0.4t$, to keep the coherence length small.}\label{softboundaryfig}
\end{figure}
\end{center}

The existence of non-topological gradient corrections to the current density means that the integrated edge current will generically evolve from a topological to non-topological value as the edge is deformed from soft to sharp.   This evolution is shown Fig.~\ref{softboundaryfig} for BdG results for a range of edge widths, using the $t'=3t/8$ lattice model, which are compared to the current predicted by (\ref{JH0}).

While the current \emph{density} is never topological near the atomically sharp edges of superconducting crystals, as noted in Sec.~\ref{preliminarysec}, there do exist special models of chiral $p$-wave superconductivity for which the \emph{integrated} current at an atomically sharp edge agrees exactly with the ``topological'' result \eqref{JH0}, valid for a soft edge.  One such model is the simple case of NN hopping and pairing on the square lattice with an edge along the $y$ direction, results for which are shown in the inset of Fig.~\ref{currentfig}(a).  In that case, the integrated current is actually independent of the length scale over which the density vanishes at the edge (unlike the case shown in Fig.~\ref{softboundaryfig}), so it maintains its topological value as we deform a soft edge into a sharp one.  

To understand this curious result,  we begin by noting a property of the energy spectrum.  For the cylindrical geometry considered in Sec.~\ref{preliminarysec} (open boundary conditions along $x$, periodic along $y$), the single-particle energy levels are enumerated by the quasi-momentum $k_y$ as well as an eigenvalue $j$ associated with the choice of potential or boundary conditions implemented along $x$. The usual particle-hole redundancy of the BdG equations is reflected as follows:  for each value of $j$ and $k_y$, there exists a $\bar{j}$ satisfying 
\beq E(k_y,j)=-E(-k_y,\bar{j}) \label{ph_sym}\eeq
This relation allows us to write the integrated current for the cylindrical geometry with NN hopping as 
\beq I_y = -\frac{1}{4\pi}\int dk_y v_y(k_y) \eta(k_y),\label{IyBdG}\eeq
where $v_y = 2t\sin k_y$ is the velocity and 
\beq \eta(k_y)\equiv \sum_j\mathrm{sgn}(E(k_y,j))\eeq
is the \emph{spectral asymmetry}~\cite{Stone87}.  
This result--valid for both soft and sharp boundary conditions along $x$--shows that the only way the \emph{total} integrated edge current (i.e., for a cylinder of width $2L$, the integrated current between $-L$ and $+L$) $I_y$ can change as one or both edges are adiabatically deformed is if there is \emph{spectral flow} of the eigenvalues across zero energy.  That is, the total current only changes if unoccupied states [$E(k_y,j)>0$] evolve to occupied ones [$E(k_y,j)<0$] or vice-versa.  

Spectral flow was invoked by Volovik~\cite{Volovik95} (see also Stone and Gaitan~\cite{Stone87}) to argue that the angular momentum of a disc of $N$ superfluid $^3$He-$A$ atoms would be equal to $N/2$ in the weak-coupling BCS limit as long as there is no spectral flow as the chemical potential is tuned from large and negative (the so-called ``BEC limit''~\cite{Mermin80,BCSBEC} where this value for the angular momentum is evident) to the Fermi energy in the BCS limit.  The absence of spectral flow in a disc geometry through this BCS--BEC crossover has been confirmed recently for continuum chiral $p$-wave superfluids in Ref.~\onlinecite{Tada14}.   

We consider instead the related crossover from a soft to sharp edge in a cylinder geometry, amounting to an evolution of the \emph{local} chemical potential $ \mu - A_0(\br)$.  Specifically,  consider the situation where both boundaries, one at $x\ll 0$ and the other at $x\gg 0$, are initially soft, such that the integrated currents between $(-L,0)$ and $(0,L)$ are both topological, given by $\pm (C/4\pi)\mu(0)$, where $\mu(0)$ is the bulk chemical potential at $x=0$.  These two currents are equal in magnitude but opposite in sign such that the total integrated current $I_y$ over $(-L,L)$ is zero.  Now imagine deforming one of the edges, say the one in the domain $x>0$,  into a sharp one.  Since the integrated current over $(-L,0)$ remains unchanged (the two edges are \emph{very} far apart), the integrated current at the sharp edge will remain equal to its soft-edge value if and only if the total current remains equal to zero.  I.e., spectral flow as an edge is deformed is required in order for the integrated current at a sharp edge to be different than that at a soft one.  In turn, since the total integrated current is initially zero, the spectral asymmetry $\eta(k_y)$ must evolve to a nonzero value.

\begin{figure}[!hr]
\centering
\includegraphics[width=0.9\linewidth]{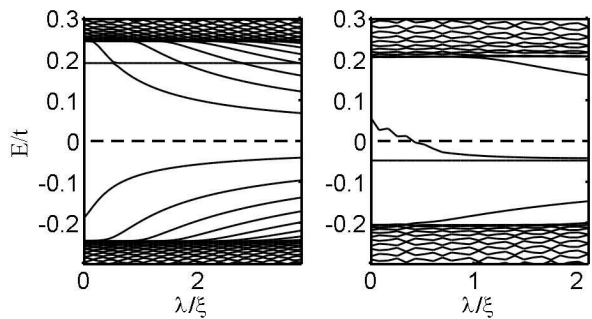}
\caption{Spectral flow plots showing the evolution of BdG eigenvalues for $E^{(j)}_{k_y=-0.29\pi}$, $\Delta(\bk) = \Delta_0(\sin k_x + i\sin k_y)$  (left) and $E^{(j)}_{k_y=-0.26\pi}$, $\Delta(\bk) = \Delta_0(\sin k_x\cos k_y + i\sin k_y\cos k_x)$ (right) for NN-only hopping and $\mu=-t$ as the edge width $\lambda$ is evolved.}
\label{spectralfig}
\end{figure}

  \begin{figure}[!hr]
\centering
\includegraphics[width=0.9\linewidth]{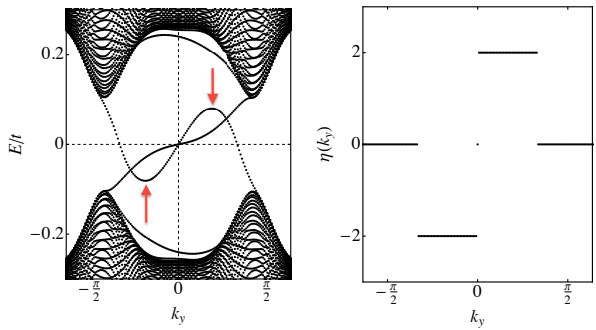}
\caption{Dispersion (left) and spectral asymmetry (right) for the NNN-pairing model  $\Delta(\bk) = \Delta_0(\sin k_x\cos k_y + i\sin k_y\cos k_x)$ for NN-only hopping at $\mu = -t$ where one edge is sharp and the other is soft.  Left: arrows point to the Majorana branch at the sharp edge.  Unlike the soft-edge branch which only crosses zero at $k_y=0$, the sharp edge branch has an additional zero-crossing away from $k_y=0$, as expected from the spectral flow shown in Fig~\ref{spectralfig}.  This extra zero-crossing gives rise to the nonzero spectral asymmetry shown in the right panel.}
\label{spectralfig2}
\end{figure}

In Fig.~\ref{spectralfig} we compare the spectral flows of the BdG spectrum for NN hopping but with 
order parameters corresponding to NN and NNN pairing as the edge is evolved from soft to sharp.  Consistent with the results in Ref.~\onlinecite{Tada14}, there is no spectral flow for NN pairing for the smoothly varying edge potentials that we consider.  This is related to the symmetry-protection of the $k_y=0$ crossing of the chiral edge branch. Particle hole redundancy (\ref{ph_sym}) is incompatible with any continuous shift up or down in energy, as would be required to have spectral flow, of the edge modes near $E=0$ and $k_y=0$.  In the finite strip geometry studied here, the two lowest energy $k_y=0$ edge modes are separated by a finite gap [which scales as $\exp(-L/\xi_0)$] and so it is clear that one cannot have spectral flow at $k_y=0$\cite{Tada14}. However, the symmmetry (\ref{ph_sym}) ensures no spectral flow even in the thermodynamic limit where this gap closes.  By contrast, zero crossings away from $k_y=0$ do not individually satisfy (\ref{ph_sym}), but come in pairs with the same chirality, such that the pair of edge modes satisfy (\ref{ph_sym}). In this case, one can continuously shift the states up or down in energy while satsifying (\ref{ph_sym}), so spectral flow is allowed.

The absence of spectral flow for the case of NN pairing explains why the edge current retains its topological value: analogous to the constancy of the angular momentum of a chiral $p$-wave superfluid through the BCS--BEC crossover, the integrated edge current does not change as the edge is deformed, and it remains equal to the topological value inferred from the gradient expansion.  The absence of spectral flow also explains why a continuum chiral $p$-wave superfluid in a disc with sharp edges~\cite{Stone04,Sauls11,Tada14} has the same total angular momentum $N/2$ as one confined to a harmonic trap, where the density vanishes slowly~\cite{Stone08}.  

In contrast, for the case of NNN pairing, shown in the lower panel of Fig.~\ref{spectralfig}, there are zeroes in the excitation spectrum at momenta $k_y\neq 0,\pi$, giving rise to spectral flow under edge deformation.  These zeroes arise not only from the additional chiral edge branches that open up when the Chern number changes, but even for lower filling fractions, as the single Majorana branch at zero momentum bends over and crosses $E=0$ elsewhere as well.  For the spectral flow shown in Fig.~\ref{spectralfig}, there is a single Majorana branch ($C=1$) and the spectral flow is due entirely to this additional zero-crossing of this branch.  In Fig.~\ref{spectralfig2}(a), we show the dispersion for the case where one edge is sharp while the other is soft.  Consistent with the spectral flow shown in Fig.~\ref{spectralfig}, the Majorana branch for the soft edge with a single zero-crossing at $k_y=0$ evolves into one with additional zero crossings at the sharp edge.  As with non-$p$-wave superfluids~\cite{Tada14},  these zeroes at $k_y\neq 0,\pi$ provide channels for spectral flow and hence, a nonzero spectral asymmetry [see Fig.~\ref{spectralfig2}(b)] and non-topological value of the integrated current moving to the sharp-edge limit.

Introducing NNN hopping, the integrated edge current can be written as a sum of (\ref{IyBdG}) and another component $I'_y$ involving the NNN velocity operator $v'_y \equiv 4t'\sin k_y f(a)$, where $a$ is some quantum number appropriate for the potential or boundary conditions implemented along $x$ and $f$ reduces to to $\cos k_x$ in the sharp-edge limit.   As the edge is deformed, $f(a)$ and hence, $I'_y$ will evolve even without spectral flow across zero energy, although the component described by (\ref{IyBdG}) will remain constant without it.  

\section{Discussion}
\label{discussionsec}

In this paper, we have reconciled the non-universal and non-topological nature of the edge currents in chiral $p$-wave superconductors, as inferred from BdG calculations~\cite{Matsumoto99,Imai12,Imai13,Lederer14}, with na\"ive expectations for a topological  value based on the leading-order Chern--Simons term in a gradient expansion of the action~\cite{Volovik88,Goryo98}.  While the integrated edge current is always dictated by a Chern number in the soft edge limit in which the density varies over a length scale much longer than the coherence length---as would happen, for instance, in a chiral $p$-wave atomic gas superfluid confined to a harmonic trap~\cite{Stone08}---non-topological gradient corrections to the current can arise outside this limit.  Using numerical BdG and quasiclassical calculations of lattice $p$-wave superconductors, we have investigated the evolution of the integrated edge current as the edge is evolved from soft to sharp.  Symmetry-allowed physics such as next-nearest-neighbor hopping and gap 
anisotropy lead to evolution away from the topological value.  

In the special cases---certain lattice models with restricted hopping as well as the continuum limit of all models---where the integrated current is found to be topological when the edge is sharp, we have shown how this result follows from the soft-edge topological value by invoking the spectral flow of the BdG eigenvalues as the edge is deformed.  In general though, the non-topological nature of the edge current in a topological superconductor means that the edge current is sensitive to effects such as band structure~\cite{Matsumoto99,Imai12,Imai13,Lederer14}, and gap anisotropy, as well as to disorder and pair-breaking surface effects~\cite{Ashby09}.  Even for a topologically trivial superconductor with zero Chern number, such as would arise in the putative chiral $p$-wave superconductor $\Sr$ were pairing only to arise on the quasi-one dimensional $\alpha$ and $\beta$ bands~\cite{Raghu10}, the edge current is generically substantial~\cite{Imai12,Imai13,Lederer14}.  

On the other hand, the non-topological nature of the edge current means that circumstances could arise in $\Sr$,~\cite{Lederer14} in which the total edge current arising from all three bands is strongly suppressed compared to predictions based on continuum systems~\cite{Matsumoto99,Furusaki01}.   For instance, even with a fixed band structure, gap anisotropy can reduce the edge current substantially below naive expectations.  If a sufficiently anisotropic gap is present in $\Sr$, this reduction, along with some interface effect~\cite{Sauls11,Imai12,Lederer14,Bouhon14}, may reconcile theory with experiment.

\acknowledgements
We thank Jim Sauls, Mike Stone, Cenke Xu, Meng Cheng and Yi-Zhuang You for helpful discussions.  This work is supported by NSERC and CIFAR
at McMaster and by the Canada Research Chair and Canada
Council Killam programs and the National Science Foundation under Grant No. NSF PHY11-25915 (CK). At Stanford (SL) this work is supported in part by the DOE Office of Basic Energy Sciences, contract DEAC02-76SF00515, and an ABB fellowship.

\appendix

\section{Topological expressions for the edge current in lattice models}
\label{topcurrentapp}

Integrating (\ref{JH0}) gives $I_y = C(\mu)[A_{0}(x_{\mathrm{edge}})-A_0(\mathrm{bulk})]/4\pi$ for the integrated edge current where the location $x_{\mathrm{edge}}$ of the edge is determined by the point where the effective local chemical potential $\mu - A_0(x)$ equals its ``vacuum value'' $\mu_{\mathrm{vac}}$.  Assuming that $A_0$ is zero in the bulk,  this result can thus be written as
\beq I_y = C(\mu)[\mu-\mu_{\mathrm{vac}}]/4\pi.\label{Iytop2}\eeq
The value of $\mu_{\mathrm{vac}}$ depends on whether the Fermi surface (FS) is hole- or electron-like. In the former case, it corresponds to the value of the chemical potential at the top of the band whereas in the latter case, it is the chemical potential at the bottom of the band. For the model with no NNN hopping, $\mu_{\mathrm{vac}} = -4t$ in the bottom half of the band ($\mu<0$) and $4t$ in the top half ($\mu>0$), and (\ref{Iytop2}) reduces to
\beq  I_y = \frac{1}{4\pi}(4t-|\mu|).\label{SRI}\eeq  

For $t' \neq 0$, the topological expression fails to describe the current at intermediate $\mu$ and so we restrict our attention to $\mu$ near the bottom and top of the band. For $t'=3t/8$, there is a van Hove singularity at $\mu = 1.5t$ where the FS changes from being electron-like to hole-like.  This change is accompanied by a change in the Chern number from $C=1$ to -1.  For the electron-like FS, the bottom of the band is $\mu_{\mathrm{vac}} = -5.5t$.  For the hole-like FS, the top of the band is $\mu_{\mathrm{vac}} = 2.5t$.  Thus, using (\ref{Iytop2}), 
\beq I_y = \begin{cases}
\frac{1}{4\pi}(5.5t+\mu) &  \text{for $\mu$ near band bottom.}\\
\frac{1}{4\pi}(2.5t - \mu)& \text{ for $\mu$ near band top.}
\end{cases}\label{Iy2}
\eeq

For $t' = t$ and $\mu<0$, there is an electron-like FS with $C=1$ and $\mu_{\mathrm{vac}}=-8t$. For $\mu>t$, there are two hole pockets, each with $C=1$ for a total Chern number of 2, and $\mu_{\mathrm{vac}} =4t$. Hence, 
\beq I_y = \begin{cases}
\frac{1}{4\pi}(8t+\mu) & \text{for $\mu$ near band bottom.} \\
\frac{2}{4\pi}(\mu-4t)& \text{for $\mu$ near band top.}
\end{cases}\label{Iy3}
\eeq

\section{Quasiclassical expression for the edge current}
\label{QCapp}
In this section we extend the calculation of the integrated edge current in Ref. \onlinecite{Sauls11} to an arbitrary band structure and either two or three spatial dimensions. We consider a single-band problem on a lattice, near an edge or (surface) parallel to a reflection plane of the bulk band structure. In three dimensions, we further assume a symmetry of the superconducting state under reflection through a horizontal plane. For triplet order parameters we assume a fixed $\bf d$-vector axis, which we take to be $\bf z$, so that the spin structure is trivial for both triplet and singlet cases.  We assume a chiral order parameter $\Delta_x(\vec p) + i \Delta_y(\vec p)$, where $\Delta_{x,y}$ are real and $\vec p$ represents a momentum vector on the Fermi surface. Furthermore, we neglect the texture of the order parameter in the vicinity of the edge, and so take $\Delta_{x,y}$ to equal their uniform bulk values. Note that the presence of a sharp edge formally invalidates the quasiclassical approximation, which is valid only on length scales much greater than $k_F^{-1}$, so the edge physics is incorporated here as a phenomenological boundary condition. Form the quasiclassical propagator

\begin{align}
\widehat {\mathcal{G}} (\vec r,\vec p; i\omega_n) = \left( \begin{array}{cc} g(\vec r,\vec p; i\omega_n)& f(\vec r,\vec p; i\omega_n) \\  f^*(\vec r,\vec p; i\omega_n) & -g(\vec r,\vec p; i\omega_n)\end{array} \right) 
\end{align}

This object is essentially the Nambu propagator integrated with respect to relative momentum. It depends on the center of mass position $\vec r$, the Fermi surface momentum $\vec p$ and the Matsubara frequency $\omega_n\equiv (2n+1)\pi T$.  $\widehat{\mathcal{G}}$ obeys the Eilenberger equation:

\begin{align}
	i \vec v \cdot \nabla_{\vec r} \widehat{ \mathcal{G}} = -\left[\widehat {\mathscr{H}},\widehat{\mathcal{G}}\right],\qquad \text{where}
\end{align}

\begin{align}
\widehat {\mathscr{H}}= \left( \begin{array}{cc} i\omega_n & \Delta_x(\vec p)-i \Delta_y(\vec p) \\    -\Delta_x(\vec p)-i \Delta_y(\vec p) & -i\omega_n\end{array} \right) 
\end{align}
and $\vec v$ is the Fermi velocity at momentum $\vec p$ (with $\vec p$ on the Fermi surface). $\widehat{\mathcal{G}}$ is taken to obey the normalization condition
\begin{align}
\label{eq:norm}
\left(\widehat{\mathcal{G}}\right)^2 = -\pi^2
\end{align}
If we decompose $\widehat{\mathcal{G}}$ into Pauli matrices according to
\begin{align}
\widehat{\mathcal{G}}=g\widehat{\tau_3} + i f_2 \widehat{\tau_1} - i f_1 \widehat{\tau_2},
\end{align}
and form the column vector
\begin{align}
| \mathcal{G} \rangle =\left(\begin{array}{c} f_1 \\f_2 \\ g\end{array}\right)
\end{align}
then $|\mathcal{G}\rangle$ obeys the (vector) differential equation
\begin{align}
\label{eq:diff}
\frac{1}{2} \vec v \cdot \nabla_{\vec r} |\mathcal{G}\rangle = \widehat{M}  |\mathcal{G}\rangle
, \qquad\text{where}\end{align}

\begin{align}
\widehat{M} = \left( \begin{array}{ccc} 0 & i\omega_n & \Delta_y \\
			-i\omega_n & 0 & - \Delta_x \\
			\Delta_y & -\Delta_x & 0 \end{array}\right)
\end{align}
Solutions to eq.  are exponential in position, with the decay length determined by the eigenvalues of $\wh{M}$. Since $\wh{M}$ is Hermitian these are real, and have eigenvectors:
\bea{}
|0;\vec p\rangle = \frac{1}{\lambda}\left(\begin{array}{c} -\Delta_x \\-\Delta_y \\ i\omega_n \end{array}\right)
\end{align}
for eigenvalue $0$, where $\lambda=\sqrt{\omega_n^2+\Delta_x^2+\Delta_y^2}$, and
\bea{}
|\pm;\vec p\rangle = &\frac{1}{\sqrt{2}\lambda\lambda_1}\left(\begin{array}{c} \pm i\omega_n \lambda-\Delta_x\Delta_y \\ \lambda_1^2  \\ i\omega_n\Delta_y \mp \lambda\Delta_x \end{array}\right),\quad
\end{align}
for eigenvalues $\pm \lambda$, where $\lambda_1=\sqrt{\omega_n^2+\Delta_x^2}$
We now assume the edge is along $x=0$, with the superconductor in the region $x>0$, and use translation invariance along $y$ to write down the generic solution of \eqref{eq:diff}:
\bea
|\G(x, \vec p)\rangle =& C_0|0; \vec p\rangle + C_+\exp\left(\frac{2\lambda }{v_x}x\right)|+;\vec p\rangle \\&+C_-\exp\left(-\frac{2\lambda }{v_x}x\right)|-;\vec p\rangle \nonumber
\end{align}
We must exclude solutions which explode as $x\rightarrow+\infty$. If we define $s\equiv \text{sign}(v_x)$, then
\bea
|\G(x, \vec p)\rangle = C_0|0; \vec p\rangle + C_{-s} \exp\left(-\frac{2\lambda }{|v_x|}x\right)|-s;\vec p\rangle 
\end{align}
The normalization condition fixes $C_0 = -\pi$. $C_{-s}$ is determined by boundary conditions at $x=0$, namely that $|\G(0,\vec p)\rangle=|\G(0,\vec{ \underline{ p}})\rangle$, where $\vec {\underline{p}}$ is the 
specular reflection of $\vec p$, i.e. $\vec{ \underline{p}}=(-p_x,p_y)$. Applying this condition yields
\bea
C_{-s}=\frac{\sqrt{2}\pi \Delta_x\lambda_1}{i\omega_n\lambda s + \Delta_x\Delta_y}
\end{align}
The current density is computed from the normal part of the propagator according to
\bea
\vec J(\vec r) = \frac{2T}{(2\pi)^d}\sum_{i\omega_n}\oint_{F.S} \frac{d\vec p}{|\vec v|}\vec v\times g(\vec r, \vec p; i\omega_n)
\end{align}
Where $d$ is the spatial dimension. The current flows solely in the $y$ direction and depends only on $x$:
\bea
J_y(x) = \frac{2T}{(2\pi)^d}\sum_{i\omega_n}\oint_{F.S} \frac{d\vec p}{|\vec v|}v_y\times g(x, \vec p; i\omega_n)
\end{align}
Our solution for the normal part of the propagator can be written 
\bea
g(x,\vec p;i\omega_n)=-&\pi \frac{i\omega_n}{\lambda}+\\ &\pi \frac{\Delta_x}{\lambda}\frac{i\omega_n\Delta_x-\lambda s\Delta_y}{(i\omega_n)^2-\Delta_y^2}\exp\left(-\frac{2\lambda}{|v_x|}x\right)\nonumber
\end{align}
The only part of the above whose contribution to the current density does not vanish by symmetry is 
\bea
\tilde g(x,\vec p; i\omega_n)=\pi\frac{s\Delta_x\Delta_y}{\omega_n^2+\Delta_y^2}\exp\left(-\frac{2\lambda}{|v_x|}x\right)
\end{align}
We compute the integrated current
\bea
I_y=&\int_0^{\infty}dxJ_y(x)\\
=&\frac{2T}{(2\pi)^d}\int_0^{\infty}dx\sum_{i\omega_n}\oint_{F.S} \frac{d\vec p}{|\vec v|}v_y\times \tilde g(x, \vec p; i\omega_n) \nonumber\\
=&\frac{1}{2(2\pi)^d}\oint_{F.S} \frac{d\vec p}{|\vec v|}v_x v_y \Delta_x\Delta_y\nonumber\\&\times 2\pi T\sum_{i\omega_n}\frac{1}{\lambda}\frac{1 }{\omega_n^2+\Delta_y^2}\nonumber
\end{align}
In the zero-temperature limit the Matsubara sum becomes an integral: $2\pi T\sum_{i\omega_n}\rightarrow\int d\omega$
\bea
I_y=&\frac{1}{2(2\pi)^d}\oint_{F.S} \frac{d\vec p}{|\vec v|}v_x v_y \Delta_x\Delta_y\times \\&\times\int_{-\infty}^{\infty}d\omega \frac{1}{\sqrt{\omega^2+\Delta_x^2+\Delta_y^2}}\frac{1}{\omega^2+\Delta_y^2}\nonumber
\end{align}
The integral has a closed form solution:
\bea
\int_{-\infty}^{\infty}d\omega \frac{1}{\sqrt{\omega^2+\Delta_x^2+\Delta_y^2}}\frac{1}{\omega^2+\Delta_y^2}\\=\frac{2}{|\Delta_x||\Delta_y|}\tan^{-1}\left(\frac{|\Delta_x|}{|\Delta_y|}\right)\nonumber
\end{align}
yielding
\bea
\label{eq:final}
I_y=&\frac{1}{(2\pi)^d}\oint_{F.S} \frac{d\vec p}{|\vec v|}v_x v_y\tan^{-1}\left(\frac{\Delta_x}{\Delta_y}\right)
\end{align}
Equation \eqref{eq:final} is the main result.  It shows that the edge current, unless prohibited by symmetry, is generically equal to the Fermi energy times a number of order one and fundamental constants. 

Up until now we have taken the order parameter to be chiral but assumed nothing about its symmetry, and very little about the point group symmetry. We now specialize to a tetragonal point group and consider various possible chiral order parameters.  For an order parameter of symmetry $d_{xy}+id_{x^2-y^2}$, the total current of equation \eqref{eq:final} vanishes by symmetry. The total current also vanishes by symmetry (under reflection $y\rightarrow -y$) for a $d_{x^2-y^2}+is$ order parameter, though the analysis above does not directly apply in that case.  

When $\Delta_{x,y}$ correspond to a two-dimensional representation of the tetragonal point group (i.e. the order parameter has either $p_x+ip_y$ or $d_{xz}+id_{yz}$ symmetry), there is a useful simplification of \eqref{eq:final}.  Under a $90$ degree rotation in the $k_xk_y$ plane, 
\bea
&\Delta_x\rightarrow\Delta_y,\quad \Delta_y\rightarrow -\Delta_x, \quad v_x\rightarrow v_y,\quad v_y\rightarrow -v_x.  \nonumber \\
&v_xv_y\tan^{-1}\left(\frac{\Delta_x}{\Delta_y}\right)\rightarrow  v_xv_y\tan^{-1}\left(\frac{\Delta_y}{\Delta_x}\right) \nonumber \\
&\quad=v_xv_y\left(\frac{\pi}{2}\text{sign}(\Delta_x\Delta_y)-\tan^{-1}\left(\frac{\Delta_x}{\Delta_y}\right)\right) \nonumber
\end{align}
Accordingly the relative magnitudes of $\Delta_{x,y}$ are unimportant, and the zero temperature current is determined only by the sign structure of the order parameter components on the Fermi surface: 
\bea
I_y=&\frac{\pi}{4(2\pi)^d}\oint_{F.S} \frac{d\vec p}{|\vec v|}v_x v_y\text{sign}(\Delta_x\Delta_y)
\end{align}
In the absence of ``accidental'' zeros of either order parameter component, the dependence on $\Delta_{x,y}$ drops out entirely, except for the overall chirality $\eta=\text{sign}(v_xv_y\Delta_x\Delta_y)$:
\bea
I_y=&\frac{\pi\eta}{4(2\pi)^d}\oint_{F.S} \frac{d\vec p}{|\vec v|}|v_x v_y|
\end{align}
For the special case of a two-dimensional system in which the dispersion relation separates according to $\epsilon(k)=\epsilon_x(k_x)+\epsilon_y(k_y)$, this simplifies to:
\bea
I_y=&\frac{\pi\eta}{(2\pi)^2}\int_{\text{first quadrant}} dp_y v_y = \frac{\eta}{4\pi}\int_{0}^{\mu} d(\epsilon_y) = \frac{\eta \mu}{4\pi}
\end{align}
Which coincides with the result gleaned from the gradient expansion, with $\eta$ equal to the Chern number.

\section{Gradient expansion of the mean-field BCS action for a chiral $p$-wave superconductor}
\label{actionapp}

The effective Euclidean Bose action for a superconductor has the usual form~\cite{Popovbook,Aitchison95}
 \beq S_{\mathrm{eff}} =-\int d^{2+1}x\int d^{2+1}x' \frac{|\Delta(x,x')|^2}{V(\br,\br')}  - \mathrm{Tr}\ln \bG^{-1},
\label{Seff} \eeq
where $V(\br,\br')$ is the attractive effective interaction  that supports $p$-wave superconductivity and $x = (\br, \tau)$ where $\tau \equiv it$ is the Wick-rotated imaginary time variable.  Consistent with mean-field BdG, we will deal with the mean-field, ``saddle-point'' value of this action by ignoring fluctuations of the phase of the order parameter.   The  inverse of the mean-field $2\times 2$ matrix Nambu-Gorkov Green's function is thus ($\hbar=e=c=1$)
\begin{widetext}
\beq \bG^{-1}(x,x')=\left(\begin{array}{cc}\left[-\partial_{\tau}-\frac{\left(-i\bnab + \bA\right)^2}{2m} + \mu +A_0\right]\delta(x-x')&\Delta(\br,\br')\delta(\tau-\tau')\nonumber\\ \Delta^{*}(\br',\br)\delta(\tau-\tau') & \left[-\partial_{\tau}+\frac{\left(i\bnab + \bA\right)^2}{2m} - \mu -A_0\right]\delta(x-x')\end{array}\right),\label{g0}\eeq
\end{widetext}
where the minimal coupling scheme $\partial_{\tau}\to \partial_{\tau}-A_0, \bnab  \to \bnab + i\bA$ has been used.

We now expand (\ref{Seff}) in gradients of the static potential
\beq A_0(\br) = A_0\sin \bQ\cdot\br\label{grad1}\eeq
as well as gradients of the order parameter amplitudes
\beq \Delta_{\alpha}(x,x') = \Delta_{0,\alpha}(\br-\br')\eta_{\alpha}\left(\frac{\br+\br'}{2}\right)\delta(\tau-\tau')\label{ansatzA}\eeq
with
\beq \eta_{\alpha}(\br) = 1+\lambda_{\alpha}\sin \bQ\cdot\br.
\label{ansatzB}\eeq
Here $\Delta_{0,\alpha}$ are the complex mean-field order parameter components for $\alpha=x,y$ (only dependent on the relative coordinate $\br-\br'$) and 
$\eta_{\alpha}$ is the corresponding amplitude, equal to unity in the absence of an external potential.   A gradient expansion need not be applied to the vector potential $\bA(\br)$ since this will be set to zero at the end of the calculation of the current and we can simply treat it as small, retaining only terms in the action that are linear in $\bA$.  

Fourier transforming (\ref{Seff}) to Matsubara frequency/momentum space $k\equiv (\bk,i\omega_n)$, the logarithm is expanded as
\begin{align}\mathrm{Tr}\ln [\bG(k,k')^{-1}] &= \mathrm{Tr}\ln [\bG_0^{-1}] +\mathrm{Tr}[(\bG_0\bSig) ] \nonumber\\ &+ \frac{1}{2}\mathrm{Tr}[(\bG_0\bSig)^2 ]+\cdots\label{expansion}\end{align}
where ($\hat{\tau}_{\alpha}$ are Pauli matrices)
\beq \bG^{-1}_0(k) = i\omega_n - \xi(\bk)\hat{\tau}_3 + \mathrm{Re}\Delta_0(\bk)\hat{\tau}_1 - \mathrm{Im}\Delta_0(\bk)\hat{\tau}_2,\label{G0}\eeq
\begin{align} &\bSig(k,k') = -\frac{1}{m}\sum_{\bq}\bA(\bq)\cdot (\bk-\bq/2)\delta_{\bk',\bk-\bq} + \Big\{\Big[ \frac{A_0}{2i}\hat{\tau}_3 \nonumber\\ &+\frac{\lambda_{\alpha}}{2i}[\mathrm{Re}\Delta_{\alpha}(\bk-\bQ/2)\hat{\tau}_1-\mathrm{Im}\Delta_{\alpha}(\bk-\bQ/2)\hat{\tau}_2]\Big]\delta_{\bk',\bk-\bQ}\nonumber\\&-(\bQ \to -\bQ)\Big\},\label{Sigma}\end{align}
and the trace is performed over frequency and momentum variables in addition to Nambu indices.  $\bSig=0$ when $\bQ=0$ and $\bA=0$ and consequently, (\ref{expansion}) constitutes a perturbative expansion in powers of $\bQ$ and $\bA$.

Using (\ref{G0}) and (\ref{Sigma}) in (\ref{expansion}), the leading order gradients terms in the action [given by the second term on the right-hand side of (\ref{expansion}) and discarding terms of order $\bA^2$] are (from hereon in, we reserve $\mu$ and $\nu$ to denote Cartesian components $x,y$)
\begin{align} S^{(2)} =&
\frac{A_0A_{\mu}(\bQ)}{2i}\chi_{0\mu}(\bQ) +\frac{\lambda_{\mu}A_{\nu}(\bQ)}{2i}\chi_{\Delta_{\mu} \nu}(\bQ)\nonumber\\&-(\bQ\to -\bQ).\label{S2k}\end{align}
Here we only show the gradient terms involving the vector potential $\bA$ since only these contribute to the current.  The following \emph{static} correlation functions have been defined [$k\equiv (\omega_n,\bk)$, and $q \equiv (0,\bQ)$, where $\omega_n$ is a Fermi Matsubara frequency and the external Bose Matsubara frequency is zero]: 
\beq \chi_{0\mu}(\bQ) \equiv \frac{1}{\beta}\sum_k v_{\mu}(\bk) \mathrm{tr}[\bG_0(k+\tfrac{q}{2})\hat{\tau}_3\bG_0(k-\tfrac{q}{2})]\label{chi0mu},\eeq
is the density-current correlation function, and, taking $\Delta_{x}(\bk)$ and $\Delta_y(\bk)$ to be purely real and imaginary, respectively, 
\begin{align} &\chi_{\Delta_{x}\nu}(\bQ) \equiv \frac{1}{\beta}\sum_k \Delta_{x}(\bk) v_{\nu}(\bk) \mathrm{tr}[\bG_0(k+\tfrac{q}{2})\bG_0(k-\tfrac{q}{2})\hat{\tau}_{1}],\nonumber\\ &\chi_{\Delta_{y}\nu}(\bQ) \equiv \nonumber\\&-\frac{1}{\beta}\sum_k \Delta_{y}(\bk) v_{\nu}(\bk) \mathrm{tr}[\bG_0(k+\tfrac{q}{2})\bG_0(k-\tfrac{q}{2})\hat{\tau}_{2}] \label{chiDmu}\end{align}
are the order parameter-current correlation functions.  $v_{\mu}\equiv \partial_{\bk_{\mu}}\xi(\bk)$ is the bare velocity vertex.

Continuing with the gradient expansion, we expand the static correlation functions (\ref{chi0mu}) and (\ref{chiDmu}) in powers of $\bQ$.  At $T=0$, 
\beq \chi_{0y}(\bQ) =-iQ_x\sum_{\bk}\frac{v_y\Delta_y(\partial_{k_x}\Delta_x)}{2E^3_{\bk}} +{\cal{O}}(\bQ^3),\label{chi0y}\eeq
\beq \chi_{0x}(\bQ) =iQ_y\sum_{\bk}\frac{v_y\Delta_y(\partial_{k_x}\Delta_x)}{2E^3_{\bk}}  +{\cal{O}}(\bQ^3),\label{chi0x}\eeq
\beq  \chi_{\Delta_x \mu}(\bQ) =-iQ_{\nu}\sum_{\bk}\frac{v_{\mu}\Delta_x}{2E^3_{\bk}}\left[v_{\nu}\Delta_y - \xi(\partial_{k_{\nu}}\Delta_y)\right] +  {\cal{O}}(\bQ^3)\label{chiDxy},\eeq
and
\beq  \chi_{\Delta_y \mu}(\bQ) =iQ_{\nu}\sum_{\bk}\frac{v_{\mu}\Delta_y}{2E^3_{\bk}}\left[v_{\nu}\Delta_x - \xi(\partial_{k_{\nu}}\Delta_x)\right] +  {\cal{O}}(\bQ^3).\label{chiDyy}\eeq
The first terms in the square brackets in (\ref{chi0y}) and (\ref{chi0x}) are both equal to the Chern number modulo particle-hole corrections:
\begin{align} &\sum_{\bk}\frac{v_y\Delta_y\left(\partial_{k_x}\Delta_x\right)}{2E^3_{\bk}} = \sum_{\bk}\frac{v_x\Delta_x\left(\partial_{k_y}\Delta_y\right)}{2E^3_{\bk}} \nonumber\\&=\frac{C}{4\pi} + {\cal{O}}(\Delta^2_0/E^2_F).\label{sigmaxy}\end{align}
Note that this is the static Hall conductivity $\tilde{\sigma}_{xy}$ defined in Section~\ref{topologyHe3sec}.  Turning to (\ref{chiDxy}) and (\ref{chiDyy}), the first term in square brackets is the $\gamma$ coefficient, also shown in (\ref{gammaxy}).  The second term in both expressions is ${\cal{O}}(\Delta^2_0/E^2_F)$ and is related to the difference in the Ginzburg--Landau coefficients $k_3$ and $k_4$~\cite{Ashby09}, which can also be obtained from the order parameter-current correlation function (\ref{chiDmu}), albeit in the limit $T\to T_c$ instead of $T=0$.   

Using the long-wavelength limiting values (\ref{chi0y})-(\ref{sigmaxy}) in (\ref{S2k}) and Fourier-transforming back to real-space gives Chern--Simons action (\ref{Leff}) plus the amplitude contribution (\ref{LD}).

The generalization of the above results to lattice models is straightforward.  As long as the coherence length $\xi_0$ is much longer than $k^{-1}_F\sim a$, where $a$ is the lattice spacing, the hydrodynamic Lagrangian retains the same form as (\ref{Leff}), with only a few minor modifications to the coefficients.  For a single-band model, one can simply use the expressions (\ref{chi0y})-(\ref{sigmaxy}) for the hydrodynamic coefficients using values appropriate for a lattice model, e.g. (\ref{xi}) and (\ref{OP}) instead of $\xi(\bk) = \bk^2/2m-\mu$ and $\Delta_0(\bk) = \Delta_0(k_x+ik_y)/k_F$.  For multiband models, one must go back and evaluate the correlation functions (\ref{chi0mu}-\ref{chiDmu}) using the appropriate higher-dimensional matrix Green's functions.



\begin{thebibliography}{99} 
\bibitem{Beenakker13} C.~W.~J.~Beenakker, \href{http://www.annualreviews.org/eprint/czvBPBNssJfFxJPueB2x/full/10.1146/annurev-conmatphys-030212-184337}{Ann. Rev. Cond. Matt.  \textbf{4}, 113} \href{http://www.annualreviews.org/eprint/czvBPBNssJfFxJPueB2x/full/10.1146/annurev-conmatphys-030212-184337}{(2013).}  
\bibitem{Niu85} Q.~Niu, D.~Thouless, and Y.-S.~Wu, \href{http://prb.aps.org/abstract/PRB/v31/i6/p3372_1}{Phys. Rev. B \textbf{31}, 3372 (1985).} 
\bibitem{Imai12} Y.~Imai, K.~Wakabayashi, and M.~Sigrist, \href{http://prb.aps.org/abstract/PRB/v85/i17/e174532}{Phys. Rev. B \textbf{85}, 174532 (2012)}.  
\bibitem{Imai13} Y.~Imai, K.~Wakabayashi, and M.~Sigrist, \href{http://prb.aps.org/abstract/PRB/v88/i14/e144503}{Phys. Rev. B \textbf{88}, 144503 (2013).}
\bibitem{Lederer14} S.~Lederer, W.~Huang, E.~Taylor, S.~Raghu, and C.~Kallin, \href{http://dx.doi.org/10.1103/PhysRevB.90.134521}{Phys. Rev. \textbf{90}, 134521 (2014).}
\bibitem{Matsumoto99} M.~Matsumoto and M.~Sigrist, \href{http://jpsj.ipap.jp/link?JPSJ/68/994/}{J. Phys. Soc. Jpn. \textbf{68}, 994 (1999).}
 \bibitem{Furusaki01} A.~Furusaki, M.~Matsumoto, and M.~Sigrist, \href{http://prb.aps.org/abstract/PRB/v64/i5/e054514}{Phys. Rev. B \textbf{64}, 054514 (2001).}  
\bibitem{Kirtley07} J.~Kirtley, C.~Kallin, C.~Hicks, E.-A.~Kim, Y.~Liu, K.~Moler, Y.~Maeno, and K.~Nelson, \href{http://prb.aps.org/abstract/PRB/v76/i1/e014526}{Phys. Rev. B \textbf{76}, 014526 (2007).} 
\bibitem{Hicks10} C.~W.~Hicks, J.~R.~Kirtley, T.~M.~Lippman, N.~C.~Koshnick, M.~E.~Huber, Y.~Maeno, W.~M.~Yuhasz, M.~B.~Maple, and K.~A.~Moler, \href{http://prb.aps.org/abstract/PRB/v81/i21/e214501}{Phys. Rev. B \textbf{81}, 214501 (2010).}
\bibitem{Jang11} J.~Jang, D.~G.~Ferguson, V.~Vakaryuk, R.~Budakian, S.~B.~Chung, P.~M.~Goldbart, Y.~Maeno, \href{http://www.sciencemag.org/content/331/6014/186.short}{Science, \textbf{331}, 186 (2011).}
\bibitem{Curran14} P. J.~Curran, S. J.~Bending, W. M.~Desoky, A. S.~Gibbs, S. L.~Lee, and A. P.~Mackenzie, \href{http://link.aps.org/doi/10. 1103/PhysRevB.89.144504}{Phys. Rev. B. \textbf{89}, 133404 (2014).}
\bibitem{Stone04} M.~Stone and R.~Roy, \href{http://prb.aps.org/abstract/PRB/v69/i18/e184511}{Phys. Rev. B \textbf{69}, 184511 (2004).}
\bibitem{Volovik88} G.~E.~Volovik, \href{http://www.jetp.ac.ru/cgi-bin/e/index/e/67/9/p1804?a=list}{Sov. Phys. JETP \textbf{67}, 1804 (1988).} 
\bibitem{Volovik92} G.~E.~Volovik, \href{http://www.jetpletters.ac.ru/ps/1273/article_19263.shtml}{JETP Lett. \textbf{55}, 368 (1992).}
\bibitem{Goryo98} J.~Goryo and K.~Ishikawa, \href{http://www.sciencedirect.com/science/article/pii/S0375960198004381}{Phys. Lett. A \textbf{246}, 549 (1998).}
\bibitem{Girvin87} S.~M.~Girvin and A.~H.~MacDonald, \href{http://prl.aps.org/abstract/PRL/v58/i12/p1252_1}{Phys. Rev. Lett. \textbf{58}, 1252 (1987).}
\bibitem{Zhang89} S.~C.~Zhang, T.~H.~Hansson, and S.~Kivelson, \href{http://prl.aps.org/abstract/PRL/v62/i1/p82_1}{Phys. Rev. Lett. \textbf{62}.}
 \bibitem{Sauls11} J.A.~Sauls, \href{http://prb.aps.org/abstract/PRB/v84/i21/e214509}{Phys. Rev. B \textbf{84}, 214509 (2011).}
\bibitem{Volovik95} G.~E.~Volovik, \href{http://www.jetpletters.ac.ru/ps/1210/article_18297.shtml}{Sov. Phys. JETP \textbf{61}, 958 (1995).} 
\bibitem{Scaffidi14} T.~Scaffidi and S.~H.~Simon, \href{http://arxiv.org/abs/1410.6073}{arXiv:1410.6073}.
 \bibitem{Mackenzie03} A.~P.~Mackenzie and Y.~Maeno, \href{http://rmp.aps.org/abstract/RMP/v75/i2/p657_1}{Rev. Mod. Phys. \textbf{75}, 657 (2003).}
\bibitem{Kallin09} C.~Kallin and A.~J.~Berlinsky,  \href{http://iopscience.iop.org/0953-8984/21/16/164210/}{J. Phys.: Condens. Matter \textbf{21}, 164210 (2009).}
\bibitem{Kallin12} C.~Kallin, \href{http://iopscience.iop.org/0034-4885/75/4/042501/}{Rep. Prog. Phys. \textbf{75}, 042501 (2012).}
\bibitem{Maeno12} Y.~Maeno, S.~Kittaka, T.~Nomura, S.~Yonezawa, and K.~Ishida, 
\href{http://jpsj.ipap.jp/link?JPSJ/81/011009/}{J.~Phys. Soc. Jpn. \textbf{81}, 011009 (2012).}
\bibitem{Roy08} R.~Roy and C.~Kallin, \href{http://prb.aps.org/abstract/PRB/v77/i17/e174513}{Phys. Rev. B \textbf{77}, 174513 (2008).} 
\bibitem{Mermin80} N.~D.~Mermin and P.~Muzikar, \href{http://prb.aps.org/abstract/PRB/v21/i3/p980_1}{Phys. Rev. B \textbf{21}, 980 (1980).}
\bibitem{Damascelli00} A.~Damascelli, D.~H.~Lu, K.~M.~Shen, N.~P.~Armitage, F.~Ronning, D.~L.~Feng, C.~Kim, Z.-X.~Shen, T.~Kimura, Y.~Tokura, Z.~Q.~Mao, and Y.~Maeno, \href{http://dx.doi.org/10.1103/PhysRevLett.85.5194 }{Phys. Rev. Lett. \textbf{85}, 5194 (2000).} 
\bibitem{Huang14} W.~Huang, E.~Taylor, and C.~Kallin, \href{http://arxiv.org/abs/1410.0377}{arXiv:1410.0377}. 
\bibitem{Ambegaokar74} V.~Ambegaokar, P.~G.~de Gennes, and D.~Rainer, \href{http://pra.aps.org/abstract/PRA/v9/i6/p2676_1}{Phys. Rev. A \textbf{9}, 2676 (1974).}
\bibitem{Ashby09} P.~Ashby and C.~Kallin, \href{http://prb.aps.org/abstract/PRB/v79/i22/e224509}{Phys. Rev. B \textbf{79}, 224509 (2009).}
\bibitem{gradientnote} In the sharp-edge limit, the short-distance physics of $A_0$ is implicitly incorporated  as phenomenological boundary conditions on the order parameter identical to those employed in Ginzburg-Landau calculations. 
\bibitem{Stone87} M.~Stone and F.~Gaitan, \href{http://dx.doi.org/10.1016/S0003-4916(87)80014-3}{Ann. Phys.  \textbf{178}, 89 (1987). } 
\bibitem{BCSBEC} A.~J.~Leggett,  in \textit{Modern Trends in the Theory of Condensed Matter}.  \href{http://link.springer.com/chapter/10.1007/BFb0120125?no-access=true}{Lecture Notes in Physics \textbf{115}, 13 (1980).}   
\bibitem{Tada14} Y.~Tada, W.~Nie, M.~Oshikawa, \href{http://arxiv.org/abs/1409.7459}{arXiv:1409.7459.}  
\bibitem{Kiatev01} A.~Kitaev, \href{http://dx.doi.org/10.1070/1063-7869/44/10S/S29}{Phys.-Usp \textbf{44}, 131 (2001).}  
\bibitem{Stone08} M.~Stone and I.~Anduaga, \href{http://dx.doi.org/10.1016/j.aop.2007.04.020}{Ann. Phys. \textbf{323}, 2 (2008).}  
\bibitem{Raghu10} S.~Raghu, A.~Kapitulnik, and S.~A.~Kivelson, \href{http://prl.aps.org/abstract/PRL/v105/i13/e136401}{Phys. Rev. Lett. \textbf{105}, 136401 (2010).}
\bibitem{Bouhon14} A.~Bouhon and M.~Sigrist,  \href{http://arxiv.org/abs/1409.1516}{arXiv:1409.1516.} 
\bibitem{Popovbook}  V.~N.~Popov, \textit{Functional Integrals and Collective Excitations} (Cambridge Univ. Press, Cambridge, 1987).  
\bibitem{Aitchison95} I.~J.~R.~Aitchison, P.~Ao, D.~J.~Thouless, and X.-M.~Zhu, \href{http://prb.aps.org/abstract/PRB/v51/i10/p6531_1}{Phys. Rev. B \textbf{51}, 6531 (1995).}

\end{thebibliography}
\end{document}